\documentclass[12pt]{article}
\usepackage{amsfonts,epsfig,latexsym,amscd,multicol,theorem,amssymb}

 \newcommand{\R}{{\mathbb{R}}}
\newcommand{\Z}{{\mathbb{Z}}} 
 
\newcommand{\I}{{\mathbb{I}}} 
\newcommand{\CP}{{\mathbb{C}}{\mathbb{P}}}

\newcommand{\beq}{\begin{equation}} \newcommand{\eeq}{\end{equation}}
\newcommand{\bea}{\begin{eqnarray}} \newcommand{\eea}{\end{eqnarray}}
\newcommand{\ra}{\rightarrow}

\newcommand{\cp}{\smallsmile} 

\newcommand{\free}[2]{{\mbox{FreeMaps}}(#1,#2)}

\theoremstyle{plain} \newtheorem{thm}{Theorem}
\newtheorem{lemma}[thm]{Lemma} \newtheorem{prop}[thm]{Proposition}
\newtheorem{cor}[thm]{Corollary} 
{\theorembodyfont{\rmfamily} \newtheorem{defn}[thm]{Definition}

}
\newcommand{\news}{\setcounter{equation}{0}}

\begin{document}

\title{Geometry and analysis in non-linear sigma models} \author{Dave
Auckly  \thanks{The first author was partially supported by  NSF grant
DMS-0204651.}  \\ Department of Mathematics, \\ Kansas State
University, \\ Manhattan, Kansas 66506, USA \\ \\ Lev Kapitanski
\thanks{The second author was partially supported by NSF grant
DMS-0436403.}              \\ Department of Mathematics, \\ University
of Miami, \\ Coral Gabels, Florida 33124, USA \\ \\ J. Martin Speight
\thanks{The third author was partially supported by EPSRC grant
GR/R66982/01.}  \\ School of Mathematics, \\ University of Leeds, \\
Leeds LS2 9JT, England}

\maketitle

\begin{abstract} 
The configuration space of a non-linear sigma model is the space of maps from one manifold to another. This paper reviews the authors' work on non-linear sigma models with target a homogeneous space. It begins with a description of the components, fundamental group, and cohomology of such configuration spaces together with the physical interpretations of these results. The topological arguments given generalize to Sobolev maps. The advantages of representing homogeneous space valued maps by flat connections are described, with applications to the homotopy theory of Sobolev maps, and minimization problems for the Skyrme and Faddeev functionals. The paper concludes with some speculation about the possiblility of using these techniques to define new invariants of manifolds.
\end{abstract}

\section{Introduction}
\label{intro}
\news

Physicists have used variational principles to describe physical
phenomena for a long time. The classical trajectories of a system are
modeled as the stationary points of an action functional. The quantum
behavior of the system may be modeled in one of two frameworks:
Lagrangian or Hamiltonian. The Lagrangian framework uses path
integrals of an exponential of the action functional. This has proved
to be a very powerful tool for physical modeling  and mathematical inspiration. The
Hamiltonian point of view is better understood from a mathematical
point of view, but there is still work to be done in this direction
\cite{simwoo}.

The first case studied via a variational principle is the motion of a
point particle on some configuration manifold. The static configurations are the
critical points of a potential energy, and the classical trajectories
are described by a non-linear system of ordinary differential
equations. These equations are the Euler-Lagrange equations of the
action functional. The second case considers fields such as the
electro-magnetic field.  A field is a function on a manifold or more
generally a section of a vector bundle over the manifold. The
classical dynamics of a field are described by a generally non-linear
system of partial differential equations (the $U(1)$ Yang-Mills
equations for electro-magnetic fields or $SU(2)\times U(1)$ Yang-Mills
equations for electro-weak fields). These equations are the
Euler-Lagrange equations of a certain functional. Quantum Field Theory
is the quantum version of this second case. Physicists now hedge their
bets by speaking of effective field theories, \cite{weinberg}. By this
they acknowledge that Quantum Field Theories produce very accurate
predictions, but are possibly not the correct theories to describe the
universe.  They argue that in certain limiting conditions one Quantum Field
Theory will accurately model physical behavior, in other limiting
conditions some other Quantum Field Theory will provide the accurate
model. It is reasonable to expect that under certain limiting
conditions fields will take values close to the critical points of
potential energy. In this situation, one would consider the basic
objects to be maps from a manifold to some non-linear space. This is
the usual origin of non-linear sigma models.

A non-linear sigma model is a model with a configuration space
consisting of maps from one manifold $M$ representing space  to some target
manifold, $N$, \cite{bmss}. Techniques from the geometry of manifolds
can shed light on the behavior of non-linear sigma models, and it is
possible that non-linear sigma models may prove to be useful tools for
the study of the geometry of manifolds. 

In this paper, we will discuss some geometric and analytical
techniques that have been useful in the study of two non-linear sigma
models. We will begin with a discussion of the topology of the
relevant configuration spaces in the first section. In the second
section, we will discuss analytical issues related to these models.

\section{Topology of configuration spaces}
\label{top}
\news

Mathematically, the challenges within the study of non-linear sigma
models arise because the target is not a linear or convex space. The
topology of the target allows one to consider different types of
constraint in an optimization problem. These constraints are an
integral feature of the model. For example, it is common to look for a
minimizer of a functional in a fixed homotopy class. If the homotopy
class is not constrained it may be that a constant map will minimize
the functional. 
 
There are in fact many different but related configuration spaces that
one could consider. We will concentrate on two general types of
configuration space: maps into Lie groups and maps into $S^2$.  Let
$M$ be a compact, oriented 3-manifold and $G$ be any Lie group.  Then
the first configuration space we consider is  the space of continuous
maps $M\ra G$ which send a chosen basepoint $x_0\in M$ to $\I\in
G$. This space is denoted by $G^M$. The second configuration space is
$(S^2)^M$.

In this section, we will give configuration spaces the compact open
topology.  In practice some Sobolev topology depending on the energy
functional is probably appropriate. The issue of checking the
algebraic topology arguments given  for classes of
Sobolev maps is interesting.  We will address this issue in the next
section. We will motivate the study of the algebraic topology of these
spaces by the physical interpretation of the non-linear sigma
model. More detailed physical and geometric interpretations of the
topological results in this section may be found in \cite{AS}.

\subsection{Components of configuration space}\label{pi0}

The Skyrme model has been considered as a model of several different
systems. In one case it is used to model nucleons: the protons and
neutrons in the center of an atom. Space is assumed to be $\R^3$ (this
is reasonable when we are modeling the nucleus of an atom). Fields are
taken to be maps into Sp$(1)$ the group of unit quaternions.  In order
to have finite energy, the gradient of such a map must have finite
$L^2$ norm. It is natural, therefore, to impose the boundary condition
that the field approach a constant value as $|\bf{ x}|\to\infty$,
where $\bf{x}$ denotes position in space. We may choose this boundary
value to be $1$ without loss of generality.  So we take our
configuration space to be Sp$(1)^{S^3}$. This is the space of base
point preserving maps from $S^3$ to Sp$(1)$. We are viewing $S^3$ as
the one point compactification of $\R^3$, the base point, infinity, is
assumed to map to the base point $1$. Notice that topologically
Sp$(1)$ is $S^3$. The homotopy classes of maps from $S^3$ to Sp$(1)$
are classified by an integer degree. The energy minimizer in the
degree $0$ class is just the constant map, in physicists' language,
the vacuum.

Linearizing the field equations about the vacuum we
obtain a wave equation for a field taking values in
$\mathfrak{sp}(1)\cong\R^3$. Travelling wave solutions of this wave
equation represent the $\pi^+$, $\pi^-$ and $\pi^0$ mesons which
transmit the strong nuclear force between nucleons. Upon quantization,
the Fourier coefficients of a general solution of the wave equation
become the pion creation and annihilation operators of the theory. The
neutron and proton appear in this model as the degree $1$ minimizers
of the energy functional. More precisely, they are the lowest energy
quantum states built on top of a degree $1$ minimizer. Deciding which
states are protons and which neutrons is a matter of convention since
the theory models only the strong nuclear force, which is insensitive
to this distinction (in physicists' language, it is isospin
invariant). 

Return to the physical interpretation of the degree.
Quantum states built on degree $B$ minimizers are taken to
represent bound states of $B$ nucleons, that is, nuclei of atomic mass
$B$.  Thinking of the degree $1$ minimizers themselves as nucleons,
the particles are smeared out over space. There are several ways to
define the center of such a particle. We could take it to be the
position of  maximum energy density, or maximum particle number
density (the latter being the inner product of the volume form on
space with the pull-back of the volume form on Sp$(1)$). One simple
and useful choice is to take the particle to be centered where its
field differs most from its boundary value. In other words, we imagine
a point in space mapped to $-1$ to be the center of a particle.  The
field may be a thought of as a smooth bump with support near such a
point. The field could also appear as several bump functions
representing several particles. There is an orientation in this
situation and the positively oriented bumps may be considered
particles and the negatively oriented bumps may be considered as
anti-particles. The degree may be computed as the number of inverse
images of a regular point. Since these images represent the nucleons
and anti-nucleons, we see that the degree $B$ is the net nucleon
number of the field. (Actually, the quantum states built on the degree
$1$ minimizer represent a class of matter particles collectively
called baryons, the higher excited states being somewhat exotic. For
this reason, the degree of a field is called its baryon number in the
Skyrme model literarture.)  The path components of the configuration
space consist of all maps with a fixed degree.  Notice that it is
possible to create or cancel particle / anti-particle pairs and stay
within the same path component. It is desirable to have a functional
that will force the trajectories to stay in a fixed path
component. This corresponds to a conservation law: there can be no net
change in the number of particles. It is exactly these considerations
that lead physicists to ask about the homotopy classification of maps
from one space to another. In the case with Lie group target we have
the following result from \cite{AK1}.

\begin{prop}\label{components}
Let $G$ be a compact Lie group and $M$ be a connected, closed
$3$-manifold.  The set of path components of $G^M$ is
$$
G^M/(G^M)_0\,\cong\, H^3(M; \pi_3(G))\times H^1(M; H_1(G)).
$$
\end{prop} 

\noindent
In the case of Lie group valued maps the configuration space has the
structure of a topological group, so the space of path components is
also a group. It would be nice to understand the group structure.  The
reason the above proposition only describes the set of path components
is that Auckly and Kapitanski only establish an exact sequence
$$
0\to H^3(M; \pi_3(G))\to G^M/(G^M)_0\to H^1(M; H_1(G))\to 0.
$$
To understand the group structure on the set of path components one
would have to understand a bit more about this sequence (e.g.\ does it
split?).  This is one of the open questions that needs to be
addressed.  There is a similar result for free maps, \cite{AK1}.

The homotopy classification of either based or free maps from a
$3$-manifold ($3$-complex in fact) to $S^2$ was worked out by
Pontrjagin many years ago, \cite{pont}.
\begin{thm} 
\label{ponthm}
Let $M$ be a closed, connected, oriented 3-manifold, and $\mu_{S^2}$
be a generator of $H^2(S^2;\Z)\cong\Z$.  To any based map $\varphi$
from $M$ to $S^2$ one may associate the cohomology  class,
$\varphi^*\mu_{S^2}\in H^2(M;\Z)$. Every second cohomology class may
be obtained from some map and any two maps with different cohomology
classes  lie in distinct path components of $(S^2)^M$. Furthermore,
the set of path  components corresponding to a cohomology class,
$\alpha\in H^2(M)$ is in  bijective correspondence with
$H^3(M)/(2\alpha\cp H^1(M))$.
\end{thm}

The physical description given before these two theorems interpreting
the degree as a count of particles and anti-particles is the basis of
the Pontrjagin-Thom construction. We will now review this
construction, because it is useful for interpreting these theorems in
general.

Here we follow the folklore maxim: think with intersection theory and
prove with cohomology.  The combination of Poincar\'e duality and the
Pontrjagin-Thom construction gives a powerful tool for visualizing
results in algebraic topology. If $W$ is an $n$-dimensional homology
manifold, Poincar\'e duality is the isomorphism, $H^k(W)\cong
H_{n-k}(W)$. It is tempting to think of the $k$-th cohomology as the
dual of the $k$-th homology. This is not far from the truth. Putting
these together, we see that every degree $k$ cohomology class
corresponds to a unique $n-k$ cycle (codimension $k$ homology cycle),
and the image of the cocycle applied to a $k$-cycle is the weighted
number of intersection points with the corresponding $n-k$-cycle. For
field coefficients this is the entire story since there is no torsion
and the $\hbox{Ext}$ group vanishes. With other coefficients, this
gives the correct answer up to torsion. The Pontrjagin-Thom
construction associates a framed codimension $k$ submanifold of $W$ to
any map $W\to S^k$. The associated submanifold is just the inverse
image of a regular point. This is well defined up to cobordism. Going
the other way, a framed submanifold produces a map $W\to S^k$ defined
via the exponential map on fibers of a tubular neighborhood of the
submanifold and as the constant map outside of the neighborhood. 

We first consider the geometric description of the components of
$G$-valued maps for the compact, simple, simply-connected Lie
groups. These are listed, along with their centre and rational cohomology
in Table \ref{tbl0}.
In this case, the $H^1(M;H_1(G))$-factor is trivial. To
interpret the $H^3(M;H_3(G))$-factor, we will use universal
coefficient duality to view elements of $H^3(M)$ as functionals on
$H_3(M)$. Using Poncar\'e duality we view $H_3(G)$ as
$H^{{\rm dim}\,G-3}(G)$. Thus, elements of $H^3(M;H_3(G))$ may be interpreted as
functions that associate an integer to a $3$-cycle in $M$ and a
codimension $3$-cycle in $G$. Any map, $u:M\to G$ generates such a
function given by the signed count of the intersection points of the
$u$-image of the $3$-cycle and the codimension $3$-cycle. There is a
different way to say this.  Fix a trivialization of the normal bundle
to the codimension $3$ cycle in $G$.  To a map $u:M\to G$ (and the
fundamental $3$-cycle $[M]$) we associate the  collection of points,
$u^{-1}(F)$. Such a point is positively oriented if the  push forward
of an oriented frame at the point has the same orientation as the
trivialization of the normal bundle at the image. Conversely, to any
finite collection of signed points we may  associate a based map,
$u:M\to G$. Using a positively or negatively oriented  frame at each
point, we construct a diffeomorphism from the closed tubular
neighborhood of  each point  to the $3$-disk of radius $\pi$ in the
space of purely imaginary quaternions, ${\mathfrak sp}(1)$. Via the
exponential map,  $\hbox{exp}:{\mathfrak sp}(1)\to \hbox{Sp}(1)$ given
by,  exp$(x)=\cos(|x|)+\frac{\sin(|x|)}{|x|}x$ we define a map from
the  closed tubular neighborhood of the points  to Sp$(1)$. This map
may be extended to the whole $3$-manifold by sending  points in the
complement of the neighborhood to $-1$. We next modify the map by
multiplying by $-1$, so that the base point will be $1$. Finally,  we
notice that, $H_3(G)$ is generated by a homomorphic image of  Sp$(1)$,
so we can complete our map into $G$ by composition with this
inclusion. We are of course assuming that $F$ is dual to the image of
this Sp$(1)$. 

\begin{table}[t]
\centering
\begin{tabular}{|l|l|l|}
\hline\noalign{\smallskip} group, $G$ & center, $Z(G)$ &
$H^*(G;{\mathbb Q})$ \\ \hline $A_n= \hbox{SU}(n+1)$, $n\ge 2$ &
${\mathbb Z}_{n+1}$  &  ${\mathbb Q}[x_3, x_5, \dots x_{2n+1}]$ \\
\hline $B_n=\hbox{Spin}(2n+1)$, $n\ge 3$ & ${\mathbb Z}_2$ &
${\mathbb Q}[x_3, x_7, \dots x_{4n-1}]$ \\ \hline $C_n=\hbox{Sp}(n)$,
$n \ge 1$ & ${\mathbb Z}_2$ &  ${\mathbb Q}[x_3, x_7, \dots x_{4n-1}]$
\\ \hline $D_n=\hbox{Spin}(2n)$, $n \ge 4$ &  ${\mathbb Z}_2\oplus
{\mathbb Z}_2$ for even $n$  &  ${\mathbb Q}[x_3,  x_7, \dots
x_{4n-5}, y_{2n-1}]$  \\ \, & ${\mathbb Z}_4$ for odd $n$  & \\
\hline $E_6$ & ${\mathbb Z}_3$ &  ${\mathbb Q}[x_3, x_9, x_{11},
x_{15}, x_{17}, x_{23}]$ \\ \hline $E_7$ & ${\mathbb Z}_2$ &
${\mathbb Q}[x_3, x_{11}, x_{15}, x_{19}, x_{23}, x_{27}, x_{35}]$ \\
\hline $E_8$ & 0 &  ${\mathbb Q}[x_3, x_{15}, x_{23}, x_{27}, x_{35},
x_{39}, x_{47}, x_{59}]$ \\ \hline $F_4$ & 0 &  ${\mathbb Q}[x_3,
x_{11}, x_{15}, x_{23}]$ \\ \hline  $G_2$ & 0 &  ${\mathbb Q}[x_3,
x_{11}]$ \\  \hline%
\end{tabular}
\caption{Simple groups
}\label{tbl0}  
\end{table}

If the group $G$ is not assumed to be simple but is simply-connected 
and compact, it is a product  of
simple compact groups, so any map will be specified by its
components, and the components of a map will be specified by the
inverse image of the appropriate codimension $3$ cycles as described
above. If the group is no longer assumed to be simply-connected we can
associate an element of $H^1(M;H_1(G_0))$ to any map $u\in G^M$. This
element maps a loop in $M$ to a loop in $G_0$ by composing with
$u$. If $u$ and $v$ are two maps that generated the same element of
$H^1(M;H_1(G_0))$, then $u^{-1}v$ will generate the trivial element,
and so lifts to a map to the universal covering group, $\tilde G$. 
The universal
covering group of a Lie group is a product of a vector space with a
compact group, so up to homotopy this is a simply-connected, compact
Lie group. The homotopy class of the lift is specified by the element
of $H^3(M;H_3(\tilde G))$ as described previously. 

\begin{figure}
\hskip120bp\epsfig{file=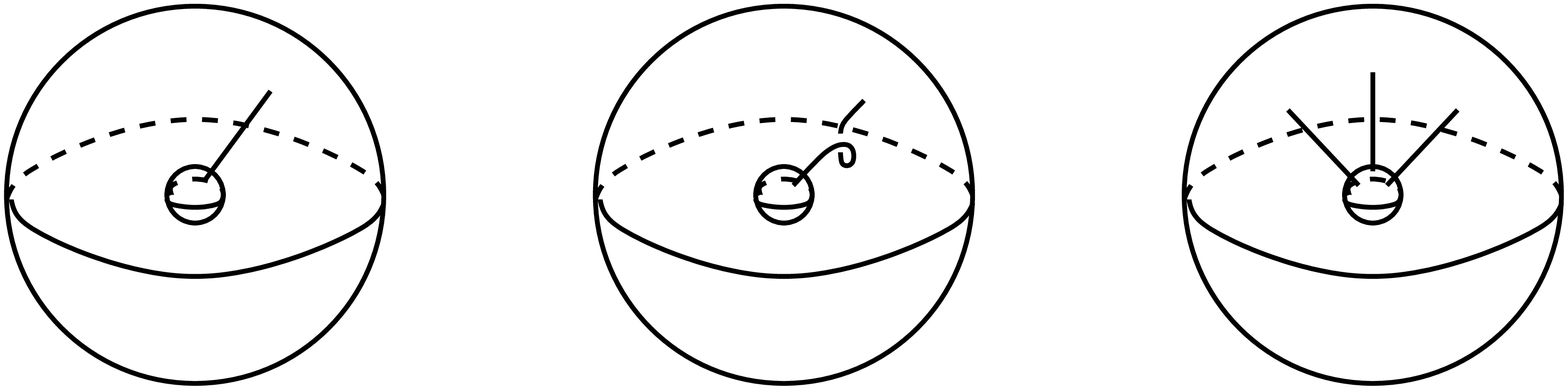,width=3truein}%
\caption{Pontrjagin-Thom representatives of $S^2$-valued
maps}\label{sputnik}
\end{figure}

Given a map $\varphi:\,M\to S^2$, denote by $(S^2)^M_\varphi$ the 
set of all maps $\psi\in (S^2)^M$ homotopic to $\varphi$. 
The picture of the components of $(S^2)^M_\varphi$ arising from the
Pontrjagin-Thom construction and Poincar\'e duality is quite nice. The
inverse image of a regular point in $S^2$ is Poincar\'e dual to
$\varphi^*\mu_{S^2}$.  This inverse image is a framed loop in
$M$. Thus such a map into $S^2$ can be viewed as modeling a spread-out
string in the same way that a map into a Lie group was viewed as
modeling a spread-out particle.  The relative number of twists in the
framing of a second map with the same pull-back is the element of
$H^3(M;\Z)/\langle 2\varphi^*\mu_{S^2}\rangle$. We will identify $S^2$
with $\CP^1$ and consider several maps to clarify the situation. We
have $\varphi_1, \varphi_1^\prime, \varphi_3:\CP^1\times S^1\to \CP^1$
given by,
$$
\varphi_1([z:w],\lambda)= [z:w],\qquad
\varphi_1^\prime([z:w],\lambda)=[\lambda z:w],\quad\hbox{and}\quad
\varphi_3([z:w],\lambda)=[z^3:w^3].
$$
We can view $\CP^1\times S^1$ as $S^2\times [0,1]$ (a spherical shell)
with the inner and outer ($S^2\times\{0\}$ and $S^2\times\{0\}$)
spheres identified. Using this convention we have displayed the framed
$1$-manifolds arising as the inverse image of a regular point in
Figure \ref{sputnik}. These figures also specify a framing. The first
vector may be taken perpendicular to the plane of the figure, and the
second vector may be obtained from the cross product with the tangent
vector.  Given two $S^2$-valued maps with the same pull-back two form,
we can arrange for the inverse image of a point in the two sphere to
be the same loop. There will be a rotation for every point on this
loop that maps the framing from the first map to the framing of the
second map. Thus any pair of maps with the same pull-back generate a
loop in SO$(2)$. The winding number of this loop is our twist number.

\begin{figure}
\hskip90bp\epsfig{file=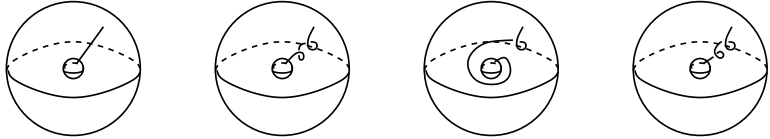,width=4truein}%
\caption{Introducing $2d$ twists}\label{untwist}
\end{figure}

It may appear that there is a well-defined twist number associated to
an $S^2$-valued map. However, there is a homeomorphism of $\CP^1\times
S^1$ twisting the $2$-sphere (such a map is given by
$([z:w],\lambda)\mapsto ([\lambda z:w],\lambda)$). This will change
the apparent number of twists in a framing, but will not change the
relative number of twists. The reason why this relative number of
twists is only well defined modulo twice the divisibility of the
cohomology class $\varphi^*\mu_{S^2}$ is demonstrated for $\varphi_1$
in Figure  \ref{untwist}.

\subsection{The fundamental group of configuration space}\label{pi1}

Where the components of the configuration space are typically labeled
by the number of particles in the system, other topological properties
of the configuration space may be used to model additional physical
properties.  We will next review the notion of statistics of particles
and describe one idea to use the fundamental group of configuration
space to incorporate statistics in a non-linear model.

Recall the distinction between bosons and fermions: a macroscopic ensemble
 of identical bosons behaves statistically as if
arbitrarily many particles can lie in the same state, while a
macroscopic ensemble of identical fermions behaves as if no two particles can
lie in the same state. Photons are examples of particles with
bosonic statistics and electrons are examples of particles with
fermionic statistics. There are several theoretical models of particle
statistics. In quantum mechanics, the wavefunction representing a
multiparticle state is symmetric under exchange of any pair of
identical bosons, and antisymmetric under exchange of any pair of
identical fermions.  In conventional perturbative quantum field
theory, commuting fields are used to represent bosons and
anti-commuting fields are  used to represent fermions. More precisely,
bosons have commuting creation operators and fermions have
anti-commuting creation operators. However, there are times when
fermions may arise within a field theory  with purely bosonic
fundamental fields. This phenomenon is called emergent fermionicity,
and it relies crucially on the topological properties of the
underlying configuration space of the model.

Emergent fermionic behavior may sometimes be explained on the basis of symmetry. Symmetries of the
classical configuration space, often imply symmetries of the space of
quantum states. However, the group of quantum symmetries can be
different from the group of classical symmetries. It is this difference that may be used to explain emergent fermionicity, \cite[chapter 7]{bmss}. A
spinning top is a well known example of this. The classical symmetry
group is SO$(3)$, while the quantum symmetry group for some
quantizations is SU$(2)$, \cite{B, sch}. An electron in the field of a
magnetic monopole is also a good example, \cite[chapter 7]{bmss} and
\cite{dirac2, z}.

When physical space is $\R^3$, so space-time is the 
usual Minkowski space,
the classical rotational symmetries induce quantum symmetries that are
representations of the $Spin$ group. These irreducible representations
are labeled by half integers (one half of the number of boxes in a
Young diagram for a representation of SU$(2)$).  The integral
representations are honest representations of the rotation group, but
the fractional ones are not. The spin statistics theorem states that
any particle with fractional spin is a fermion, and any particle with
integral spin is a boson.

The statistics of a particle may be viewed as a parity. A compound
particle made out of an even number of fermions will be a boson, and a
compound particle comprised of an odd number of fermions will be a
fermion. With this in mind, consider the fundamental group of the
space of based maps from $S^3$ to Sp$(1)$.

\begin{figure}
\hskip35bp\epsfig{file=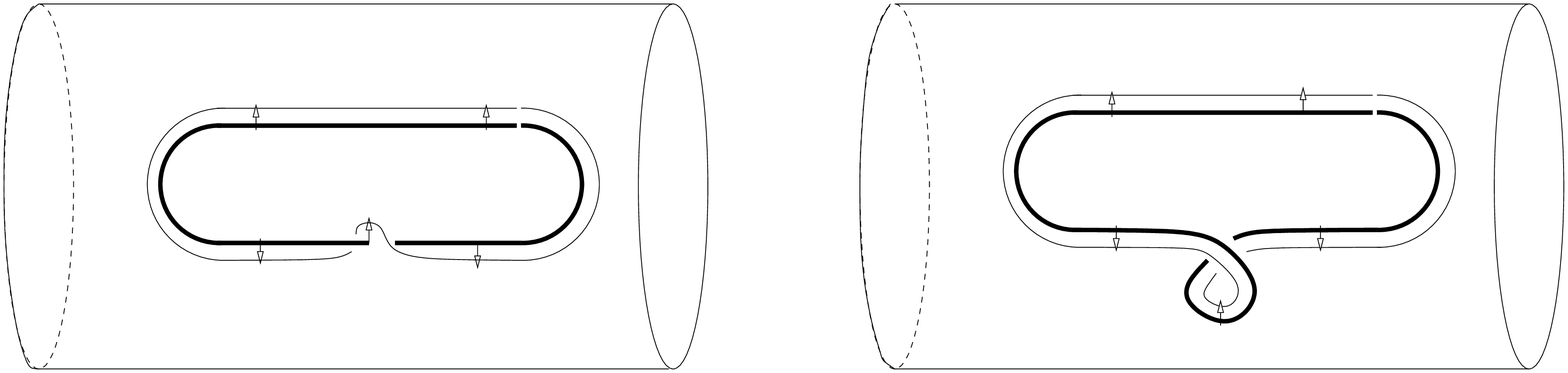,width=4truein}%
\caption{The exchange loop}\label{fig1}
\end{figure}

The following figures show some loops in this configuration space. A
loop in Sp$(1)^{S^3}$ is represented by a map from $S^3\times [0,1]$
to Sp$(1)$. Such a map may be described by a framed loop in
Sp$(1)^{S^3}$. Fibers of the closed tubular neighborhood are mapped to
Sp$(1)$ via the exponential map, and the complement of this
neighborhood can be mapped to the base point. This is exactly
analogous to the correspondence between signed points in $S^3$ and
maps into Sp$(1)$ described in the subsection on the components of Lie
group valued-maps.  The horizontal direction in these figures
represents the interval direction  in $S^3\times [0,1]$. The disks
represent the $x-y$ plane and we suppress the $z$ direction. Figure \ref{fig1} 
shows two copies of a typical loop
representing an  element in the fundamental group. Only the first
vector  of the framing is shown in Figure \ref{fig1}. The second
vector is obtained by taking the cross product with the tangent vector
to the curve in the displayed slice, and the final vector is the
$z$-direction.  It is easy to see that the  left copy may be deformed
into the right copy. We describe the left copy as  follows: a particle
and antiparticle are born; the particle undergoes a full rotation; the
two particles then annihilate.  The right copy may be described  as
follows: a first particle-antiparticle pair is born; a second pair is
born; the two particles exchange positions without rotating; the first
particle and  second antiparticle annihilate; the remaining pair
annihilates. Notice that  there are two ways a pair of particles can
exchange positions. Representing  the particles by people in a room,
the two people may step sideways  forwards/backwards and sideways
following diametrically opposite points on a  circle always facing the
back of the room. This is the exchange without  rotating described in
Figure \ref{fig1}. This exchange is non-trivial in
$\pi_1(\hbox{Sp}(1)^{S^3})$. The second way a pair of people may
change  positions is to walk around a circle at diametrically opposite
points  always facing the direction that they walk to end up facing
the opposite  direction that they started. This second change of
position is actually  homotopically trivial. Since the framed links in
Figure \ref{fig1}  avoid the slices, $S^3\times \{0, 1\}$, they
represent a loop based at the  constant identity map.

\begin{figure}
\hskip105bp\epsfig{file=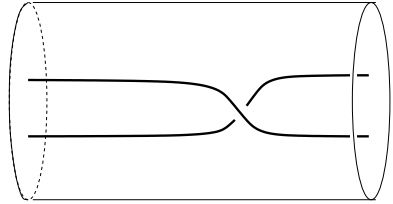,width=2truein}%
\caption{The degree $2$ exchange loop}
\label{fig2}
\end{figure}

It is possible to describe a framing without drawing any normal
vectors at  all. The first vector may be taken perpendicular to the
plane of the figure,  the second vector may be obtained from the cross
product with the tangent  vector, and the third vector may be taken to
be the suppressed $z$-direction.  The framing obtained by following
this convention is called the black board  framing. We use the
blackboard framing in figure \ref{fig2}. The  Pontrjagin-Thom
construction may also be used to visualize loops in other  components
of the configuration space. Figure \ref{fig2} shows a loop in the
degree $2$ component of the space of maps from $S^3$ to Sp$(1)$.

In fact, the exchange loop that we just described in
$\pi_1(\hbox{Sp}(1)^{S^3})$ has order two. This may be seen as
follows.  Wrapping twice around the exchange  loop is the boundary of
a disk in the configuration space. Since  ${\mathbb R}P^2$ is the
result of identifying the points on the boundary of a  disk via a
degree $2$ map, one expects to find an ${\mathbb R}P^2$ embedded  into
configuration space. In \cite{sorkin}, R.~Sorkin  describes an
embedding, $f_{{\rm stat}}:{\mathbb R}P^2\to \hbox{Sp}(1)^{S^3}$.  We
briefly recall Sorkin's elegant construction. Describe ${\mathbb
R}P^2$ as the $2$-sphere with antipodal points identified. By the
addition of particle  antiparticle pairs, we may assume that there are
two particles  in a representative map. We may place the particles at
antipodal points of a  sphere using frames parallel to the coordinate
directions. The loop in configuration space represented by a compound
particle making a full rotation is trivial in the fundamental group if
the compound particle consists of an even number of basic particles,
and is non-trivial if the compound particle consists of an odd number
of basic particles. This observation is the basic motivation behind
Finkelstein and Rubinstein's method of incorporating fermionic
solitons in the Skyrme model \cite{finrub}. The fundamental groups of
the Lie group valued and $S^2$-valued configuration spaces were
computed in \cite{AS}. The results are repeated below.

\begin{thm}\label{thm1} If $M$ is a closed, connected, orientable 
$3$-manifold, and $G$ is any Lie group, then 
$$
\pi_1(G^M)\, \cong\,{\mathbb Z}_2^s\,\oplus H^2(M; \pi_3(G)).
$$
Here $s$ is the number of symplectic factors in the lie algebra of $G$.
\end{thm}

\begin{thm}\label{thm2}  Let $M$ be closed, connected and orientable. 
For any $\varphi\in (S^2)^M$, the fundamental group of
$(S^2)^M_\varphi$ is given by
$$
\pi_1((S^2)^M_\varphi)\cong {\mathbb Z}_2\oplus H^2(M;{\mathbb
Z})\oplus  \hbox{\rm ker}(2\varphi^*\mu_{S^2}\cp).
$$
Here $2\varphi^*\mu_{S^2}\cp:H^1(M;{\mathbb Z})\to  H^3(M;{\mathbb
Z})$ is the usual map given by the cup product. 
\end{thm}

The Pontrjagin-Thom construction may also be used to understand the
isomorphism,
$$
\phi:\pi_1(G^M)\, \cong\,{\mathbb Z}_2^s\,\oplus H^2(M; \pi_3(G)),
$$
asserted in Theorem \ref{thm1}. A loop in $(G^M)_0$ based at the
constant map $u(x)=\I$, may be  regarded as a based map $\gamma:SM\to
G$. The identifications in the  suspension provide a particularly nice
way to summarize all of the constraints on $\gamma$ imposed by the
base points. We will use the same notation for the  map 
$\gamma:M\times [0,1]\to G$ obtained from $\gamma$ by  composition
with the natural projection.  The inverse image  $\gamma^{-1}(F)$ with
framing obtained by pulling  back the trivialization of $\nu(F)$ may
be associated to $\gamma$. Conversely,  given a framed link in
$(M-p_0)\times (0,1)$ one may construct an element  of
$\pi_1(G^M)$. Using the framing each fiber of the  closed tubular
neighborhood to the link may be identified with the disk of radius
$\pi$ in ${\mathfrak sp}(1)$. As before $-1$ times the  exponential
map may be used to construct  a map, $\gamma:SM\to G$ representing an
element of  $\pi_1(G^M)$.

It is now possible to describe the geometric content of the
isomorphism in  Theorem \ref{thm1}. For a class of loops $[\gamma]\in
(G^M)_0$, let $\phi(\gamma)=(\phi_1(\gamma),\phi_2(\gamma))$. Restrict
attention to the case of simply-connected $G$, and make the
identifications, $\pi_3(G)\cong  H_3(G;{\mathbb Z})\cong
H^{\hbox{dim}\,G-3}(G;{\mathbb Z})$. An element of  $H^2(M; \pi_3(G))$ may be
interpreted as a function that associates an integer to a surface in
$M$, say $\Sigma$, and a codimension $3$ cycle in $G$, say  $F$. Set
$\phi_2(\gamma)(\Sigma,F)=\#(\Sigma\times [0,1]\cap\gamma^{-1}(F))$.
Note that $\gamma^{-1}(F)$ inherits an orientation from the framing
and  orientation on $M$. Using Poincar\'e duality this may be said in
a  different way. The homology class of $\gamma^{-1}(F)$ in
$(M-p_0)\times (0,1)$ projects to an element of $H_1(M)$ dual to the
element associated to $\phi_2(\gamma)$. The first component of the
isomorphism counts the parity of the number of twists in the framing. 

Similarly we can describe the content of Theorem \ref{thm2}. An
element of $\pi_1((S^2)^M_\varphi)$ is represented by a map,
$\gamma:M\times S^1\to S^2$. The inverse image of a regular point is a
$2$-dimensional submanifold, say $\Sigma$. This defines an element of
$H^1(M;\Z)$ as follows. To any $1$-cycle in $M$, say $\sigma$, we
associate the intersection number of $\Sigma$ and $\sigma\times
S^1$. Since our loop is in the path component of $\varphi$, the
surface $\Sigma$ is parallel to the $\varphi$-inverse image of a
regular point. This implies that our element of $H^1(M;\Z)$ is in the
kernel of the map, $2\varphi^*\mu_{S^2}:H^1(M;\Z)\to H^3(M;\Z)$. Given
any element of this kernel, we can define a loop in $(S^2)^M_\varphi$
via what we call the ${\mathfrak q}$-map. This is the map  
 $\,\mathfrak q:\,S^2\times S^1\to \hbox{Sp}(1)\,$ defined by ${\mathfrak q}(x{\bf i}x^*,\lambda)=x\lambda x^*$. One can to check that every element of $S^2$ can be written in the form $x{\bf i}x^*$, and the expression $x\lambda x^*$is independent of the choice of $x$ representing the element of $S^2$.
There is
a map $u:M\times S^1\to \hbox{Sp}(1)$ that may be used to change
this new loop back into $\gamma$. The remaining homotopy invariants of
$\gamma$ are just those of $u$ as described previously.

\begin{figure}
\hskip35bp\epsfig{file=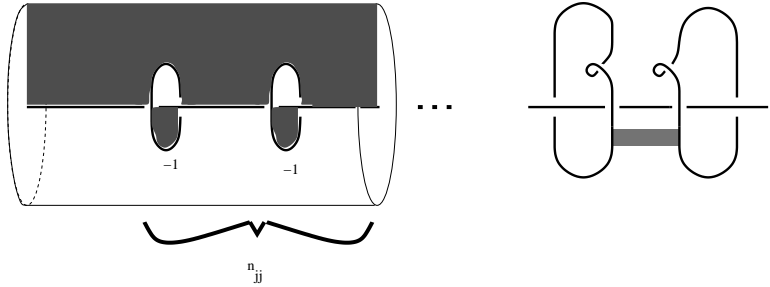,width=4truein}%
\caption{The $2$-cycle}\label{fig4}
\end{figure}

In order to convey some feel for the proofs of these results we repeat
the proof of one of the main steps from \cite{AS}.
\begin{prop}\label{split}
The sequence,
$$
0\to \pi_4(G)\to\pi_1(G^M)\to H^2(M;\pi_3(G))\to 0
$$
splits, and there is a splitting associated to each spin structure on
$M$.
\end{prop}
{\it Proof:\, } It is  sufficient to check the result for
$G=\hbox{Sp}(1)$. Since the three  dimensional Spin cobordism group is
trivial, every $3$-manifold is surgery  on a framed link with even
self-linking numbers \cite{kirby}.  Such a surgery description induces
a Spin structure in $M$. Let $M=S^3_L$ be  such a surgery description,
orient the link and let $\{\mu_j\}_{j=1}^c$ be the positively oriented
meridians to the components of the link. These meridians  generate
$H_1(M)\cong H^2(M;\pi_3(\hbox{Sp}(1)))$. This last isomorphism is
Poincar\'e duality. Define a splitting by:
$$
s:H_1(M)\to\pi_1((\hbox{Sp}(1))^M); \quad s(\mu_j)=
PT(\mu_j\times\{\frac12\}, \hbox{canonical framing}).
$$
Here $PT$ represents the Pontrjagin-Thom construction and the
canonical  framing is constructed as follows. The first vector is
chosen to be the  $0$-framing on $\mu_j$ considered as an unknot in
$S^3$. The second vector is  obtained by taking the cross product of
the tangent vector with the first  vector, and the third vector is
just the direction of the interval. We will  now check that this map
respects the relations in $H_1(M)$. Let  $Q_L=(n_{jk})$ be the linking
matrix so that,  $H_1(M)=\langle \mu_j| n_{jk}\mu_j=0 \rangle$. We are
using the summation  convention in this description. The $2$-cycle
representing the relation,  $n_{jk}\mu_j=0$ may be constructed from a
Seifert surface to the  $j^{\hbox{th}}$ component of the link, when
this component is viewed  as a knot in $S^3$. Let $\Sigma_j$ denote
this Seifert surface. The desired  $2$-cycle is then
$\widehat\Sigma_j=(\Sigma_j-\stackrel{\circ}{N}(L))\cup\sigma_j$. Here
$\sigma_j$ is the surface in $S^1\times D^2$ with $n_{jj}$ meridians
depicted  on the left in figure \ref{fig4}.
The boundary of $\widehat\Sigma_j$ is exactly the relation,
$n_{jk}\mu_j=0$.  The framing on each copy of $\mu_k$ for $k\neq j$
induced from this surface  agrees with the $0$-framing. The framing on
each copy of $\mu_j$ is  $-\hbox{sign}(n_{jj})$. The surface,
$\widehat\Sigma_j$ may be extended to a  surface in $M\times
[0,1]\times [0,1]$ by adding a collar of the boundary in  the
direction of the second interval followed by one band for each pair of
the $\mu_j$ as depicted on the right of Figure \ref{fig4}. The
resulting  surface has a canonical framing, and the corresponding
homotopy given by  the Pontrjagin-Thom construction homotopes the loop
corresponding to the  relation to a loop corresponding to a
$\pm2$-framed unlink. Such a loop is  null-homotopic, as
required. \hfill $\Box$

It is interesting to notice that spin structures enter into the
analysis of the topology of these configuration
spaces. This should not be surprising since spin structures were first introduced to model the spin behavior of point particles. The fundamental group of configuration space may be used to model the spin behavior of solitons.  $\hbox{Spin}^c$ structures enter at other points of the story.

\subsection{Rational cohomology of configuration space}

The second cohomology of configuration space plays an important role
in the quantum theory. The full rational cohomology may also appear as
topological observables in the theory. To see the role of the second
cohomology, recall some of the theory of geometric quantization
\cite{simwoo}. In quantum theory one wishes to include the algebra of
observables into the collection of operators on some function space.
Given a configuration manifold (space of positions), one defines the
phase space (collection of positions and momenta) to be the cotangent
bundle of configuration space. This has a canonical symplectic
structure.  The space of wave functions is chosen to be a set of
sections of a complex line bundle over the quotient of phase space by
an integrable isotropic distribution of maximal dimension, the
so-called polarization. The most common choice is the vertical
polarization, consisting of the tangent spaces to the fibres of the
cotangent bundle.  In this case, the quotient is naturally identified
with the configuration space itself. 

There are many different complex line bundles over a space. Line
bundles are in one to one correspondence with the elements of the
second cohomology with integral coefficients. This is called
quantization ambiguity in the physics literature.  If we choose the
vertical polarization, computing the quantization ambiguity requires
one to consider the second integral cohomology of configuration space.
It is well-known that the second integral cohomology group may be
determined from the fundamental group and the second rational
cohomology group. 

The real  cohomology ring $H^*((G^M)_0,\R)$, including its
multiplicative structure is described in the following theorem from
\cite{AS}. To state the theorem we will use a $\mu$-map.

Similar to Yang-Mills theory, there is a $\mu$-map, 
$$
\mu:H_d(M;{\mathbb R})\otimes H^j(G;{\mathbb R}) \to
H^{j-d}(G^M;{\mathbb R}),
$$
and the cohomology ring is generated (as an algebra) by the images of
this map.  The $\mu$ map produces a $(j-d)$-cocycle in $G^M$ from a
$d$-cycle in $M$ and a $j$-cocycle in $G$.  On the level of chains,
let $e^d:D^d\to M$ be a $d$-cell, and $x_j$ be a  closed $j$-form on
$G$. Given a singular simplex, $u:\Delta^{j-d}\to G^M$, let $\widehat
u:M\times \Delta^{j-d}\to G$ be the natural map and write
$$
\mu(e^d\otimes x_j)(u)= \int_{D^d\times \Delta^{j-d}} \widehat u^* x_j.
$$

\begin{thm}\label{thm1co} Let $G$ be a compact, simply connected, simple Lie group. 
The cohomology ring of any of these groups is a free
graded-commutative  unital algebra over $\R$ generated by degree $k$
elements $x_k$ for certain  values of $k$ (and with at most one
exception at most one generator for any  given degree). The values of
$k$ depend on the group and are listed in  Table \ref{tbl0} in section
\ref{pi0}. Let $M$ be a closed, connected,  orientable
$3$-manifold. The cohomology ring $H^*((G^M)_0;\R)$ is the free
graded-commutative unital algebra over $\R$ generated by the elements
$\mu(\Sigma_j^d\otimes x_k)$,  where $\Sigma_j^d$ form a basis for
$H_d(M;\R)$ for $d>0$ and $k-d>0$. 
\end{thm}
\noindent
We also have the rational cohomology of $(S^2)^M$ from \cite{AS}.
\begin{thm}\label{thm2co}
Let $M$ be closed, connected and orientable, let $\varphi:M\to S^2$,
let $\Sigma_j^d$ form a basis for $H_d(M;\R)$ for $d<3$, and let
$\{\alpha_k\}$ for a basis for
$\hbox{ker}(2\varphi^*\mu_{S^2}\cp:H^1(M;\Z)\to H^3(M;\Z))$. The
cohomology ring $H^*((S^2)^M_\varphi;\R)$ is the free
graded-commutative unital algebra over $\R$ generated by the elements
$\alpha_k$ and $\mu(\Sigma_j^d\otimes x)$,  where 
$x\in H^3(\hbox{Sp}(1);\Z)$
is the orientation class. The classes $\alpha_k$ have degree $1$ and
$\mu(\Sigma_j^d\otimes x)$ have degree $3-d$.
\end{thm}

It is interesting to compare the configuration space of SU$(2)$
connections modulo gauge
transformations to the configuration space of Sp$(1)$-valued
maps.  

\vskip.2in

\begin{center}
\begin{small}
\begin{tabular}{|c|c|c|}
\hline 
\ & Yang-Mills configuration space & Skyrme configuration space \\
\hline
input space & a $4$-manifold, $X$ & a $3$-manifold, $M$   
 \\
\hline
homotopy type & Maps$(X,B\hbox{SU}(2))$ & 
Maps$(M,\hbox{SU}(2))$ \\
\hline
integer component labels & second Chern class, $c_2$ & 
Chern-Simons invariant, cs$=Tc_2$ \\
\hline
cohomology generators & 
$\mu(\Sigma^d\otimes c_2)\in H^{4-d}({\mathcal C})$ &  $\mu(\Sigma^d\otimes Tc_2)\in H^{3-d}({\mathcal C})$ \\
\hline%
\end{tabular}
\end{small}
\vskip.2in
{Yang-Mills vs Skyrme}
\end{center}
\noindent
In Yang-Mills theory there is a
family of moduli spaces that may be used to define a family of
homology cycles in the configuration space.  Evaluating the generators
of the cohomology of the configuration space on these homology classes
produces a $4$-manifold invariant. This invariant is called the
Donaldson polynomial. It is a polynomial on the homology of the input
$4$-manifold.

It is natural to ask if there is a similar family of homology cycles
in the Skyrme configuration space. If there is such a family, it would
be possible to define a polynomial invariant of $3$-manifolds
analogous to the Donaldson polynomial.

\subsection{Other topological results}

Using standard results from algebraic topology, it is possible to
generalize the results on the fundamental group and rational
cohomology to cases where the manifold is not connected, or to
arbitrary Lie groups. It is also possible to address configuration
spaces of free maps. Details may be found in \cite{AS}. The two most
subtle results in this direction are listed below.

\begin{thm}\label{freefun}
We have $\pi_1(\free{M}{S^2}_\varphi)\cong {\mathbb Z}_2 \oplus
\left(H^2(M;{\mathbb Z})/ \langle 2\varphi^*\mu_{S^2}\rangle
\right)\oplus  \hbox{\rm ker}(2\varphi^*\mu_{S^2}\cp)$. 
\end{thm}

\begin{thm}\label{freeco}
There is a spectral sequence with $E_2^{p,q}=H^p(S^2;\R)\otimes
H^q((S^2)^M_\varphi;\R)$ converging to
$H^*(\free{M}{S^2}_\varphi;\R)$. The second differential is given by
$d_2\mu(\Sigma^{(2)}\otimes x)=2\varphi^*\mu_{S^2}[\Sigma]\mu_{S^2}$
with $d_2$ of any other generator trivial. All higher differentials
are trivial as well.
\end{thm}

\section{Functionals and analytical techniques}
\label{analysis}
\news

There are many different function spaces that may be defined for maps
from one manifold to a second manifold. 
For an analyst, 
geometry and physics are good guides to the best questions in
analysis. 
Because of the structure of the energy functionals in physics, 
it is convenient to use the language of 
Sobolev spaces. For scalar functions on a compact manifold, 
the Sobolev spaces can be defined by first localizing 
on charts by means of a partition of unity, and then using the 
description of the space for functions on $\mathbb R^n$. 
This generalizes easily to vector-valued functions. 
To define a Sobolev space of functions taking values in a manifold
(say $N$), one takes an isometric embedding of $N$ into a vector space
and considers those elements of the Sobolev space taking values in the
vector space that lie in $N$ almost everywhere. We denote by 
$W^{s,p}(M,\,N)$ the Sobolev space of maps from $M$ to $N$ 
which have $s$ derivatives in $L^p$. 

\subsection{Local representation of functions by flat connections}

One idea that has proven to be useful in the study of non-linear
models has been a local representation of functions by flat
connections. This first shows up when the codomain is a Lie
group. Given a smooth map $u:M\to G$ one can construct the Lie algebra
valued $1$-form $A=u^{-1}du$. This form satisfies $dA+A\wedge A=0$ and
can be interpreted as a flat connection. Conversely, given a flat
connection defined on a cube $I^n$ one may define a function $u:I^n\to
G$ by solving the parallel transport equation
$du(\dot\gamma(t))=u(\gamma(t))A(\dot\gamma(t))$ with $u(x_0)=I$. It
follows from the fact that $A$ is flat that the value $u(\gamma(1))$
does not depend on the path connecting $x_0$ to $\gamma(1)$. Changing
the initial value or base point will change the function $u$ by a
constant multiple. The same result holds for $L^2$ connections, 
\cite{AK1}, but
the proof is much more involved. The result from \cite{AK1} reads:

\begin{thm}
Given any $\,L^2\,$ $\,\mathfrak g$-valued $1$-form $\,A\,$ on
$\,I^m\,$  such that 
\begin{equation}\label{flat1}
dA\,+\,\frac{1}{2}\,[A,\,A]\,=\,0
\end{equation} 
in the sense of distributions, there exists  $\,u\in
W^{1,2}(I^m,\,G)\,$ such that $\,u^{-1}\in W^{1,2}(I^m,\,G)\,$  and
$\,A\,=\,u^{-1}\,du$.  Furthermore, for any two such maps, $\,u\,$ and
$\,v$, there exists  $\,g\in G\,$ so that  $\,u(x)\,=\,g\cdot v(x)$,
for almost every $\,x\in I^m$.
\end{thm}

The second local representation theorem is proved in \cite{AK3} for
$S^2$-valued maps. It reads:

\begin{thm}
If $\varphi$ is in $W^{1,3}(I^3,S^2)$ or $W^{1,2}_E(I^3,S^2)$ then
there is a  $u\in W^{1,3}(I^3,\hbox{Sp}(1))$ or $u\in
W^{1,2}(I^3,\hbox{Sp}(1))$ respectively so that $\varphi=u^*\,{\bf i}\,u$.
In either case $\hbox{Re}((u^*du)^3)\in L^1$. Furthermore, for any two
such maps $u$ and $v$ there is a map $\lambda$ in $W^{1,3}(I^3,S^1)$
or $W^{1,2}(I^3,S^1)$ respectively, so that $v=\lambda u$. 
\end{thm}

\noindent
The space $W^{1,2}_E(I^3,S^2)$ was defined to analyze the Faddeev
model. We will recall its definition in a subsection below. The map
Sp$(1)\to S^2$ given by $u\mapsto u^*{\bf i}u$ is a convenient way to
express the quotient projection Sp$(1)\to \hbox{Sp}(1)/S^1$. More
generally, if $H$ is a closed subgroup of a Lie group $G$ then
functions from $I^n$ to $G/H$ may be factored through $G$. This will
appear in future work. There is an important class of sigma models in
physics called coset models where the target is a homogenous space
$G/H$. The $G$-valued map may then be represented by a flat connection
as before.

It is possible that the same ideas could be used to represent
functions taking values in an arbitrary manifold, $N$. The
diffeomorphisms on any manifold act transitively. This means that
given any pair of points $x_0$, $x_1$ there is a diffeomorphism with
$f(x_0)=x_1$. Let Diff$(N)$ denote the topological group of all
diffeomorphisms, and let $\hbox{Stab}_{x_0}$ denote the closed
subgroup consisting of all diffeomorphisms fixing $x_0$. It is not
difficult to see that there is a homeomorphism
Diff$(N)/\hbox{Stab}_{x_0}\cong N$. Thus we could try the same idea,
factor a map into $N$ through Diff$(N)$ and then represent a
Diff$(N)$-valued map by a flat connection. The cost in this type of
representation is that the structure group is infinite dimensional in
this general case. Such issues have been addressed in gauge theory, so
it is worth remembering this possibility, because it may someday find
application. 

\subsection{Homotopy theory for finite energy maps}

Our first application of these local representation theorems will be
to the study of homotopy theory in classes of maps with finite energy. 
This question naturally arises because one common feature 
among many sigma
models is the restriction to a fixed homotopy class. 
The homotopy classes divide the particles into different sectors, 
and numerical invariants distinguishing the classes (sectors) 
bear some definite physical meaning (depending on the model). 
Good physical models describe interesting physics  
and have energy functionals that respect geometry of the 
configuration space. 

There has been a fair amount of work done
relating homotopy theory to Sobolev spaces $W^{s,p}(M,N)$, 
see \cite{Brezis} for a review of recent results. 
A paraphrase of the result of White \cite{White} gives a good
idea of what to expect: The homotopy classes of maps with one
derivative in $L^p$ restricted to the $k$-skeleton of a manifold agree
with the homotopy classes of smooth maps provided $k$ is less than
$p$. For example, the second cohomology of a space is completely
determined by the homotopy classes of maps from the $3$-skeleton of
the space into a $K({\mathbb Z},2)$ (or ${\mathbb C}P^N$ for $N$
sufficiently large). Thus one would expect the pull-back of a second
cohomology class under a $W^{1,3}$ map to be well defined. In general,
the third cohomology of a space depends on the $4$-skeleton, of course
if the space is a $3$-manifold, then the $3$-skeleton will suffice and
the pull-back will be well defined for $W^{1,3}$ maps. Similarly one
would expect the Hopf invariant to be well defined for $W^{1,3}$
maps.

In this paper we consider two energy functionals corresponding to the 
Skyrme and Faddeev models.  
The Skyrme functional for maps $u:M\to G$ is defined to be
\begin{equation}\label{skyrme}
E(u)=\int_{M} \frac12 |u^{-1}\,d u|^2\, +\,\frac14 |u^{-1}\,d u\wedge
u^{-1}\,d u|^2\;d\,\hbox{vol}_M\,.
\end{equation}
Skyrme originally considered this in the case of maps from $\R^3$ to
SU$(2)$, \cite{sky}. We described the physical interpretation of 
these maps in the section on topology.  

The Faddeev functional is defined for maps $\,\psi:\,M\to S^2$ 
as follows, \cite{Faddeev}:
\begin{equation}\label{faddeev}
E(\psi)\,=\,\int_M |d\psi|^2\,+\,|d\psi\wedge d\psi|^2\,d\hbox{vol}\,.
\end{equation}
It is related to the Skyrme functional: viewing $S^2$ as an equator 
in the $3$-sphere SU$(2)$, 
the Faddeev functional is
just the restriction of the SU$(2)$-Skyrme functional to 
$S^2$-valued maps
$\,\psi(x)\,=\,\psi^1(x)\,{\bf i} + \psi^2(x)\,{\bf j} +
\psi^3(x)\,{\bf k}$. 

For these two models, the finite energy maps are those 
$u\in W^{1,2}(M, G)$ or $\psi \in W^{1,2}(M, S^2)$ for which 
$E(u)$ or $E(\psi)$ are finite. We denote such classes of maps 
by $W^{1,2}_E$.

It is useful to keep several explicit maps in mind when considering
the homotopy theory of maps in generalized function spaces. First
consider the map from the $3$-disk to the $2$-sphere $\eta_1:D^3\to
S^2$ given by, $\eta_1(x)=x/|x|$. Simple integration suffices to
verify that $\eta_1$ is in $W^{1,p}$ for $p<3$, but is not in
$W^{1,2}_E$ or $W^{1,3}$. Notice that the integral of the
$\eta_1$-pull-back of the normalized area form on $S^2$ integrates to
$1$ on the boundary of $D^3$. Thus the pull-back on second cohomology
may not be reasonably defined for such maps. Even more regularity is
required to define the pull-back on the third cohomology. The
composition of projection of $S^3$ to $D^3$ with this $\eta_1$ map may
be patched into any smooth map from a $3$-manifold to obtain a similar
example. Similarly the function, $\eta_2:D^3\to S^1$ given by
$\eta_2(x)=\cos(\ln|x|){\bf i}+\sin(\ln|x|){\bf j}$ is in $W^{1,p}_E$
for $p<3$ but not in $W^{1,3}$, and the function $\eta_3:D^3\to S^1$
given by $\eta_3(x)=\cos(\ln|\ln|x/e||){\bf i}+\sin(\ln|\ln|x/e||){\bf
j}$ is in $W^{1,3}$ but is not continuous. These two last functions
may also be patched into maps from an arbitrary
$3$-manifold. Furthermore they may be composed with maps into any
non-trivial compact Lie group or into $S^2$. 
 
Our applications of the local representation theorems to homotopy
problems are similar to the definition of \v{C}ech cohomology. As our
first example, consider the space of flat connections modulo gauge
equivalence. It is well known that the path components are labeled by
the holonomy representation. 

It is possible to define the holonomy for $L^2_{loc}$ distributionally
flat connections. Recall from the sample functions listed above that
the local developing maps for such a connection can look like Swiss
cheese. Here is a sketch of the definition.  Given a sufficiently fine
cover of a manifold by cubes, the nerve of the cover will be homotopy
equivalent to the manifold. (The nerve is a complex with one vertex
for each cube, an edge for each non-empty intersection, triangle for
each non-empty triple intersection etc.) The edge path group of a
complex is an analogue of the fundamental group of a topological
space. It is isomorphic to the fundamental group of the topological
realization of the complex. Given an  $L^2_{loc}$ distributionally
flat connection, say $A$, we can find local developing maps $u_p$ such
that $A=u^{-1}_pdu_p$ on each cube by the local representation
theorem. Furthermore, each edge of the nerve may be labeled by the
group element relating the two local developing maps. This gives a
representation of the edge path group. The actual definition includes
the transition functions of the bundle supporting the connection, and
is only valid for central bundles. It is an open question to
generalize the definition of holonomy to arbitrary flat
connections, see \cite{AK1}.

Using the notion of holonomy, it is possible to generalize the local
representation theorem to a global representation theorem.

\begin{lemma}\label{gaugeflat}
Two $\,L^2_{loc}\,$ distributionally  flat connections on a central
bundle are gauge equivalent if  and only if they have the same
holonomy.
\end{lemma}

\noindent The homotopy invariant of a map from $M$ to $G$ living in
$H^1(M;H_1(G_0))$ may be interpreted as the holonomy of a flat
connection, and therefore is well-defined for $W^{1,2}$ maps.

There are similar applications of the local representation theorem for
$S^2$-valued maps. Given a smooth map $\varphi:M\to S^2$ one may
construct the pull-back of the orientation class $\mu_{S^2}\in
H^2(S^2;\Z)$. This class $\varphi^*\mu_{S^2}\in H^2(M;\Z)$ is also the
obstruction to lifting the map $\varphi$ to a map $u:M\to
\hbox{Sp}(1)$ such that $\varphi=u{\bf i}u^*$.  Using the \v{C}ech
picture and the local representation theorem, one may define the
obstruction class for $W^{1,3}$ or $W^{1,2}_E$ maps. Using this
obstruction we can prove the following global representation
theorem, see \cite{AK2,AK3}.

\begin{prop}\label{lift}
If $\varphi$ and $\psi$ are two maps in $W^{1,3}(M,S^2)$ or $W^{1,2}_E(M,S^2)$
then there is a $w$ in $W^{1,3}(M,\hbox{Sp}(1))$ or $w\in
W^{1,2}(M,\hbox{Sp}(1))$ respectively with $\int_M
\hbox{Re}((w^*dw)^3)<\infty$ and $\psi=w\varphi w^*$ if and only if
$\varphi^*\mu_{S^2}=\psi^*\mu_{S^2}$. If $w_1$ and $w_2$ are two such
maps, then there is a $\lambda$ in $W^{1,3}(M,S^1)$ or in
$W^{1,2}(M,S^1)$ respectively with $w_1={\mathfrak q}w_2$.
\end{prop}

\noindent The proof of this lifting theorem is similar to the proof of
local gauge slices in gauge theory. The geometry suggests a natural
system of differential equations that may be solved to give the lift.

Using de Rham theory, some homotopy invariants may be represented as
integrals of differential forms. When an invariant is integral, this
is the first thing anyone would try. Invariants living in the top
cohomology are usually represented in this way. Sometimes a homotopy
invariant takes values in a group with torsion. There are two ways we
address torsion. We use the \v{C}ech picture as we described above, or
we lift the map to a different space and realize the torsion invariant
of the original map as an integral invariant of the lift. Torsion is
introduced in this second picture when there is more than one possible
lift. In any such application of de Rham theory there is a new issue
that arises. The de Rham cohomology is cohomology with real
coefficients and the integral is the cap product. Thus {\it a priori}
such integrals are only known to be real numbers. It is possible to
use the local representation theorems to prove that these integrals
take integral values, even for Sobolev maps.  Some sample theorems
from \cite{AK3} are listed below.

\begin{thm}\label{intthm1} Let $G$ be a compact, simply-connected Lie group, let
$M$ be a closed oriented $3$-manifold, let $\Theta$ be a smooth form 
representing an integral class in $H^3(G)$, and let $u$ be a 
map in $W^{1,3}(M,G)$  or in
$W^{1,2}_E(M,G)$. Then the number $\int_M u^*\Theta$ is
an integer.
\end{thm}

It follows that the $H^3(M;H_3(\tilde G))$ invariants of homotopy
classes of maps from $M$ to $G$ are integers for Sobolev maps and
general compact Lie groups. By lemma \ref{gaugeflat}, any $L^2_{loc}$
distributionally flat connection is gauge equivalent to a smooth
connection. The Chern-Simons invariants of these two connections are
related by the degree of the gauge transformation, so we have the
following corollary.

\begin{cor}
If $A$ is a finite energy distributionally flat $L^2$ connection on a
central bundle then there is a smooth connection on the same bundle
with the same holonomy and Chern-Simons invariant.
\end{cor}

This integrality proposition combines with the lifting theorem for
$S^2$-valued maps to give a version of the secondary invariant for
Sobolev $S^2$-valued maps. We need a second integrality result to know
that the degree of the lift changes by multiples of the right amount
when the lift is changed. This is the content of the next result.

\begin{prop}\label{intprop2}
When $\psi$ and $\lambda$ are in $W^{1,3}$ or when $\psi$ is in
$W^{1,2}_E$ and $\lambda$ is in $W^{1,2}$, the expression
$-\frac{1}{16\pi^2 i}\int_M\psi d\psi\wedge d\psi \lambda^*d\lambda$
is an integer.
\end{prop}

\noindent Here is the formal definition of the secondary homotopy
invariant for Sobolev $S^2$-valued maps.

\begin{defn}
Given $\varphi, \psi:M\to S^1$ in $W^{1,3}$ or in $W^{1,2}_E$ 
such that
$\varphi^*\mu_{S^2}=\psi^*\mu_{S^2}$ define 
$$
\Upsilon(\varphi,\psi)=
-\frac{1}{12\pi^2}\int_M\hbox{Re}((w^*dw)^3)\in {\mathbb Z}_{2m_\psi},
$$
where $w$ is the map given by Proposition \ref{lift}.
\end{defn}

It follows from Theorem \ref{intthm1}  that 
$\Upsilon$ is an integer. To see that $\Upsilon$ is well defined
notice that
$$
-\frac{1}{12\pi^2}\int_M\hbox{Re}(((w{\mathfrak
q}(\psi,\lambda)^*dw)^3) =  -\frac{1}{8\pi^2 i}\int_M\psi d\psi\wedge
d\psi \lambda^*d\lambda -\frac{1}{12\pi^2}\int_M\hbox{Re}((w^*dw)^3).
$$
See \cite{AK2} for a proof of this.  These invariants serve to
generalize Pontrjagin's theorem to Sobolev maps or finite energy maps
because two smooth $S^2$-valued maps $\varphi$ and $\psi$ are
homotopic if and only if $\varphi^*\mu_{S^2}=\psi^*\mu_{S^2}$ and
$\Upsilon(\varphi,\psi)=0$.

Our expression for $\Upsilon$ generalizes an integral formula for the
Hopf invariant, see \cite{Bott-Tu}. Indeed, 
when $\varphi^*\mu_{S^2}$ is torsion (in
particular if $M=S^3$), the pull-back, $\varphi^*\omega_{S^2}$,  of the volume form on $S^2$ is exact and there is a $1$-form, 
$\theta^\varphi$, so that
$d\theta^\varphi=\varphi^*\omega_{S^2}$. The Hopf invariant of
$\varphi$ is then defined to be $\int_M\theta^\varphi
d\theta^\varphi$. The following proposition relates 
the Hopf invariant and $\Upsilon$.

\begin{prop}
Let $\varphi, \psi:M\to S^2$ in $W^{1,3}$ or in 
$W^{1,2}_E$ such that
$R\;\varphi^*\mu_{S^2}=0$ for some non zero integer. For the minimal positive $R$,  we have
$$
\hbox{\rm Hopf}\,(\varphi)=\frac{1}{R} \Upsilon(\varphi_R,{\bf i}),
$$
where $\varphi_R$ is the composition of $\varphi$ with the 
map $z\mapsto z^R$ of $S^2$ to itself.
\end{prop}
\noindent Thus we see that our generalization of Pontrjagin's  theorem
for Sobolev maps specializes to an integrality result for the Hopf
invariant of Sobolev maps.

\subsection{Minimizing functionals}

One motivation for our study of the topology of spaces of maps and
extensions to Sobolev maps was to consider minimization problems 
for the Skyrme and Faddeev functionals. In physical lingo, the 
minimizers of the energy functional (in every sector) are 
called the ground states. Their geometric structure and stability 
are of great importance. The reader may find interesting pictures 
and discussion  
of purported ground states for the original Skyrme model in 
\cite{GP} and for the Faddeev model in \cite{Faddeev-Niemi} 
and \cite{HS-2}.  

In this subsection we will present 
our results establishing the existence of  minimizers and
compactness of the set of minimizers for the Skyrme and Faddeev 
functionals. Along with the energy functionals for maps, 
(\ref{skyrme}) and (\ref{faddeev}), we consider functionals 
for connections. In \cite{sky}, Skyrme 
noticed and used the fact that maps could
be represented by flat connections, $a=u^{-1} du$. This leads to a second version of
the Skyrme functional,
\begin{equation}\label{a-skyrme}
E[a]=\int_{M} \frac12 |a|^2\,+\,\frac1{16}
|[a,\,a]|^2\;d\,\hbox{vol}_M\,.
\end{equation}
The Faddeev functional can be also written in 
terms of flat connections, \cite{AK2}. 
  Given a smooth reference map  $\,\varphi :\,M\to S^2$, any
$\,\psi\,$ homotopic to $\,\varphi$,  can be represented as
$\,\psi\,=\,u\,\varphi\,u^*\,$ with  $\,u:\,M\to\hbox{Sp}(1)$. Plugging
this expression into the energy functional,  using
$\,\hbox{Ad}$-invariance of the norm and the Lie bracket, and the
notation  $\,a = u^*\,du\,$, we obtain
\begin{equation}\label{a-faddeev}
E(\psi)\,=\,E_{\varphi}[a]\,:=\,\int_M |D_a \varphi|^2\,+\, |D_a
\varphi\wedge D_a \varphi|^2\,,
\end{equation}
where $\,D_a \varphi\,=\,d\varphi\, +\,[a,\varphi]$.  
As was observed in \cite{LK,AK1,AK2}, the variational problems 
for maps and for connections are, in general, different. 
The way we approach these variational problems 
requires several steps, \cite{AK1,AK2}. 
First, we describe the homotopy classes 
for smooth maps and analytically define expressions to label 
different classes. Second, we show that those  
expressions for labels of the components 
extend to finite energy Sobolev maps (or connections). 
Third, we show that there exist minimizers in every sector 
of finite energy maps 
specified by given values of the label. We also prove in 
\cite{AK3} that the values these labels take on finite
energy configurations agree with the possible values on smooth
configurations.

There is an open question remaining related to the components of
the spaces $W^{1,2}_E$: The set of finite energy maps 
(or connections) with a
fixed label is not known to be path connected.  
So we call such sets sectors. We expect that the sectors are
the components of the spaces of finite energy maps when a suitable
topology is used.

The existence of minimizers is established by the direct method 
of the calculus
of variations, so given a sequence of maps in a fixed sector with
energy approaching the infimum, there is a subsequence converging to a
minimizer in the same sector. In particular, any sequence of minimizers
in a fixed sector contains a convergent subsequence, hence 
the set of minimizers of the Skyrme or Faddeev functional in a fixed
sector is  compact in $W^{1,2}_E$.
\medskip

Future work will address the existence of minimizers of a generalized
Skyrme functional for maps into general homogenous spaces. We 
also plan  to study the 
regularity. It is common for minimizers of a functional in a class of
functions to be more regular than generic members of the class. 
We expect to see
applications of the local representation results for functions to the
regularity of minimizers.

One may speculate that the minimizers of the Skyrme functional may form
a cycle in the configuration space of maps from $M$ to $G$. Given such
a cycle one could evaluate cohomology classes from the configuration
space to obtain topological invariants. We believe that this is
slightly naïve because the Skyrme functional lacks an important property shared by 
functional that do serve to define invariants of manifolds. Recall that sufficiently regular minimizers
of functionals similar to the Skyrme functional satisfy second order
Euler-Lagrange equations. When a functional is topologically
saturated, the minimizers are characterized as the solutions of a
first order system of equations. For example, the anti-self dual
condition in Yang-Mills theory, the pseudoholomorphic condition in
Gromov-Witten theory, or Sieberg-Witten equations in Seiberg-Witten
theory are all examples of first order systems characterizing minima.
One can write out what the first order equations corresponding to the Skyrme functional 
should be. However, these equations are overdetermined and essentially have no solutions. 
The equations are analogous to what one would have by considering the Seiberg-Witten equations
for a fixed connection and attempting to solve for the spin field. 
We believe that a suitable modification of the Skyrme functional incorporating new fields
(perhaps gauge bosons?) will be topologically saturated and
may lead to new smooth invariants of manifolds.


\begin{thebibliography}{xx}





\bibitem{AK1}  Auckly, D., Kapitanski, L.:  Holonomy and Skyrme's
 Model.  {\it Commun. Math. Phys.} \textbf{240}, 97--122 (2003) 

\bibitem{AK2}  Auckly, D., Kapitanski, L.:  Analysis of the Faddeev
 Model.  {\it Commun. Math. Phys.} accepted. (http://arxiv.org/abs/math-ph/040302)

\bibitem{AK3}  Auckly, D., Kapitanski, L.:  Integrality of invariants
 for Sobolev maps.  In progress.

\bibitem{AS} Auckly, D., Speight, J.M.: Fermionic quantization and
configuration spaces for the Skyrme and Faddeev-Hopf models.  preprint. (http://arxiv.org/abs/hep-th/0411010)

\bibitem{bmss} Balachandran, A., Marmo, G., Skagerstam, B., Stern, A.:
\textit{Classical topology and quantum states}.  New Jersey : World
Scientific, 1991


\bibitem{B} Bopp, F., Haag, Z.: Uber die moglichkeit von spinmodellen. {\it Zeitschrift fur Naturforschung} \textbf{5a},
644 (1950)

\bibitem{Bott-Tu} Bott, R., Tu, L.W.: 
\textit{Differential forms in algebraic topology}.
New York : Springer-Verlag, 1982


\bibitem{Brezis}
Brezis, H.:
The interplay between analysis and topology in some nonlinear PDE problems. 
{\it Bull. Amer. Math. Soc. (N.S.)} \textbf{40}, no. 2, 179--201 
(2003)



\bibitem{dirac2}  Dirac, P.: The theory of magnetic poles.  {\it
Phys. Rev.} {\bf 74}, 817--830 (1948)

\bibitem{Faddeev}
Faddeev, L. D.: Quantization of solitons.
Preprint IAS print-75-QS70 (1975)


\bibitem{Faddeev2}
Faddeev, L. D.:
Knotted solitons and their physical applications.
Phil. Trans. R. Soc. Lond. {\bf A} \textbf{359}, 1399--1403 (2001)


\bibitem{Faddeev-Niemi}
Faddeev, L. D., Niemi, A. J.:
Stable knot-like structures in classical field theory.
Nature \textbf{387}, 58--61  (1997)

\bibitem{fed} Federer, H.: A study of function spaces by spectral
sequences.  {\it Ann. of Math.} {\bf 61}, 340-361 (1956)
 
\bibitem{finrub} Finkelstein, D., Rubunstein,J,: Connection between
spin statisitics and kinks.  {\it J. Math.\ Phys.} {\bf 9}, 1762--1779
(1968)



\bibitem{GP}
Gisiger, T.,  Paranjape, M. B.: 
Recent mathematical developments in the Skyrme model.  
Phys. Rep. \textbf{306},  no. 3, 109--211 (1998)



\bibitem{giu} Giulini, D.: On the possibility of spinorial
quantization in the Skyrme model.  {\it Mod.\ Phys.\ Lett.} {\bf A8},
1917--1924 (1993)




\bibitem{HS-2} Hietarinta J., Salo P.: 
Ground state in the Faddeev-Skyrme model. 
Phys. Rev. \textbf{D 62}, 081701(R) (2000)


\bibitem{LK}  Kapitanski, L.: On Skyrme's model,  in:
\textit{Nonlinear Problems in Mathematical Physics and 
Related Topics II:
In Honor of Professor O. A. Ladyzhenskaya\/}, Birman et al., eds.
Kluwer, 2002, pp.229-242

\bibitem{kirby} Kirby, R.C.: \textit{The topology of $4$-manifolds}.
Lecture Notes in Mathematics 1374, New York : Springer-Verlag, 1985


\bibitem{pont} Pontrjagin, L.: A classification of mappings of the
$3$-dimensional complex into the  $2$-dimensional sphere.  {\it
Mat. Sbornik N.S.} \textbf{9:51}, 331-363 (1941) 


\bibitem{sch} Schulman, L.: A path integral for spin. {\it Phys. Rev.} 
{\bf 176:5}, 1558-1569
(1968))

\bibitem{simwoo} Simms, D.J., Woodhouse, N.M.J.: {\it Geometric
quantization}.  Lecture Notes in Physics 53, Berlin: Springer-Verlag,
1977

\bibitem{sky} Skyrme, T.H.R.: A unified field theory of mesons and
baryons.  {\it Nucl.\ Phys.} {\bf 31}, 555--569 (1962)

\bibitem{sorkin}  Sorkin, R.:  A general relation between
 kink-exchange and kink-rotation.  {\it Commun. Math. Phys.}
 \textbf{115}, 421--434 (1988) 


\bibitem{VK} Vakulenko, A. F., Kapitanski, L.:
On $S^2$-nonlinear $\sigma$-model.
Dokl. Acad. Nauk SSSR \textbf{248}, 810--814 (1979)

\bibitem{weinberg} Weinberg, S.: {\it The quantum theory of fields}. Cambridge :  Cambridge University Press, 1995

\bibitem{White}
White, B.: Homotopy classes in Sobolev spaces and the existence of energy minimizing maps,
Acta Mathematica \textbf{160}, 1-17 (1988)



\bibitem{wit} Witten, E.: Current algebra, baryons and quark
confinement.  {\it Nucl.\ Phys.} {\bf B233}, 433--444 (1983)

\bibitem{z} Zaccaria, F., Sudarshan, E., Nilsson, J., Mukunda, N.,
Marmo, G., Balachandran, A.: Universal unfolding of Hamiltonian
systems: from symplectic structures to fibre bundles. {\it Phys. Rev.}
{\bf D27}, 2327-2340 (1983)

\end{thebibliography}
\end{document}